\documentclass[a4paper,preprint,aps,super,nofootinbib]{revtex4-1}
\usepackage[margin=0.8in]{geometry}
\usepackage{graphicx}
\usepackage{textcomp}
\usepackage{xfrac}
\usepackage{pdfpages}

\usepackage[english]{babel}
\newcommand{\reffig}[1]{Fig. \ref{#1}}

\newcommand{\unit}{\ \mathrm}
\newcommand{\func}[2]{\mathrm{#1} \left( #2 \right)}

\renewcommand{\sin}[1]{\func{sin}{#1}}

\newcommand{\ket}[1]{|#1\rangle}
\newcommand{\neffsingle}{1.863 + 0.012i}

\newcommand{\neffsymmsmall}{2.036 + 0.020i}
\newcommand{\neffantisymmsmall}{1.841 + 0.010i}

\newcommand{\neffsymmlarge}{1.961 + 0.016i}
\newcommand{\neffantisymmlarge}{1.842 + 0.011i}

\newcommand{\simmode}[1]{
\begin{center}
	\begin{tabular}{c c c}
		\mbox{
		  \includegraphics{#1_Hx}
		} &
		\mbox{
		  \includegraphics{#1_Hy}
		} &
		\mbox{
		  \includegraphics{#1_H2}
		}
		\\
		\bf{(a)} & \bf{(b)} & \bf{(c)}
		\\
		\mbox{
		  \includegraphics{#1_Ex}
		} &
		\mbox{
		  \includegraphics{#1_Ey}
		} &
		\mbox{
		  \includegraphics{#1_E2}
		} \\
		\bf{(d)} & \bf{(e)} & \bf{(f)}
	\end{tabular}
\end{center}
\vspace{-0.5cm}
}

\begin{document}


# Quantum Interference in Plasmonic Circuits

*Reinier W. Heeres[1,*], Leo P. Kouwenhoven[1] and Valery Zwiller[1]*

[1]Kavli Institute of Nanoscience, Delft University of Technology,

P.O. Box 5046, 2600 GA Delft, The Netherlands

[*]Corresponding author: reinier@heeres.eu

July 2013

**Surface plasmon polaritons (plasmons) are a combination of light and a collective oscillation of the free electron plasma at metal-dielectric interfaces[1]. This interaction allows sub-wavelength confinement of light, beyond the diffraction limit inherent to dielectric structures[2]. The resulting electromagnetic fields are more intense and the strength of optical interactions between metallic structures and light-sources[3,4,5,6] or detectors[7,8] can be increased. Plasmons maintain non-classical photon statistics[9,10] and preserve entanglement on plasmon-assisted transmission through thin, patterned metallic films[11,12] or weakly confining waveguides[13]. For quantum applications[3,14] it is essential that plasmons behave as indistinguishable quantum particles. Here we report on a quantum interference experiment in a nanoscale plasmonic circuit consisting of an on-chip plasmon beam splitter with integrated superconducting single-photon detectors[15] to allow efficient single plasmon detection[16]. We demonstrate quantum mechanical interaction between pairs of indistinguishable plasmons by observing Hong-Ou-Mandel interference[17], a hallmark non-classical effect which is the basis of linear optics-based quantum computation[18]. Our work shows that it is feasible to shrink quantum optical experiments to the nanoscale and demonstrates a promising route for sub-wavelength quantum optical networks.**

In general, a 50/50 beam splitter transforms two input modes ($a_1$ and $a_2$) into two output modes ($b_1$ and $b_2$) in such a way that each input is divided equally over the two outputs, i.e. a single particle in either input has a 50% chance to end up in mode $b_1$ and a 50% chance to end up in mode $b_2$. When simultaneously providing a single particle in both input modes, in principle each of them travels independently and four outcomes are possible with equal probability. However, in quantum mechanics these options have to be added coherently, and the rather unexpected result is that for indistinguishable bosons[19] a two-particle interference effect gives only two possible outcomes: either both particles end up in mode $b_1$ or both in $b_2$, an effect known as Hong-Ou-Mandel (HOM) interference. Note that contrary to classical interference, the intensity, or average number of particles, in either output is not affected and it is just the correlation statistics that change. In terms of quantum operators the beam splitter can be described by the transformations $\hat{a}_1 \rightarrow \sqrt{T}\hat{b}_1 + i\sqrt{R}\hat{b}_2$ and $\hat{a}_2 \rightarrow i\sqrt{R}\hat{b}_1 + \sqrt{T}\hat{b}_2$, where R and T denote the reflection and transmission coefficients, and the factor $i$ corresponds to a $\pi/2$ phase-shift. Creating a particle simultaneously in modes $a_1$ and $a_2$ results in output state $2i\sqrt{(RT)}(|2\rangle_{b1}|0\rangle_{b2} - |0\rangle_{b1}|2\rangle_{b2}) + (T - R)|1\rangle_{b1}|1\rangle_{b2}$, from which it follows that for an ideal 50/50 beam splitter the $|1\rangle_{b1}|1\rangle_{b2}$ coincidence term disappears. Although a plasmonic beam splitter is not lossless, it has been shown theoretically that the interference visibility should remain $2RT/(R^2 + T^2)$ as long as the proper beam splitter phase relations are maintained[20].

The conventional approach for an optical quantum interference measurement is depicted in Fig. 1a. Photons from a photon-pair source are collected in optical fibers and, after a fiber beam splitter, detected by fiber-coupled silicon avalanche photodiodes (APDs). Alternatively, the photons can be interfered on a bulk dielectric beam splitter and detected using free-space detectors[17]. In both cases, the complete system is macroscopically large. Dielectric-based integrated optics approaches have managed to shrink such experiments considerably[21,22], but are always limited by diffraction and still large compared to current electronic components. Plasmonics uniquely offers a larger degree of confinement and therefore allows to create structures smaller than the diffraction limit. Our quantum interference scheme in Fig. 1b scales the whole experiment, except for the source, down by several orders of magnitude. In contrast to previous studies[23,24], our beam splitters are based on directional

couplers using the strongly confined waveguide plasmon modes to reach sub-wavelength length scales.

We use nanofabrication techniques to make our plasmonic beam splitters and integrated on-chip detectors[16]. This approach is flexible and allows to design plasmonic networks with parameters obtained from numerical simulations (see Supplementary Information). An SEM-image of a typical device is shown in Fig. 2a, together with a zoom-in of the waveguide coupling region and the detector area. The Superconducting Single Photon Detectors (SSPDs) consist of a meandering line (100 nm width) of a thin (~5 nm) superconducting material (niobium nitride). These devices are operated at 4 K, well below the critical temperature of ~12 K, by supplying a bias current close to the critical current (~22 µA). After absorption of a single photon the detector switches locally to the normal, resistive state[15]. The current is shunted to an amplifier and a voltage pulse can be measured. These detectors are easy to operate, sensitive over a broad wavelength range (ultraviolet – near infrared) and have a very small timing jitter (down to ~50 ps) and dead-time (~2 ns).

Optical measurements are performed in the setup shown in Fig. 2c. We first characterize the plasmon beam splitter by focusing a laser ($\lambda$ = 1064 nm) to a small spot (~1 µm) and raster-scanning this across the device while recording the number of detection events in both detectors simultaneously (Fig. 3a). Plasmon excitation is polarization sensitive and most efficient with polarization parallel to the waveguide (Fig. S2). For the scan we use a linearly polarized beam which couples to both inputs equally. The resulting photo-response maps in Fig. 3b and 3c have two very pronounced features. First of all, the detectors click when directly illuminated due to photon absorption, resulting in the left detector appearing in Fig. 3b and the right detector in Fig. 3c. Second, the two spots in the lower parts of Fig. 3b and 3c indicate that when illuminating those positions both detectors give detection events at an almost equal rate. This signal is caused by coupling into propagating plasmon modes at the end of a single input waveguide. The plasmons are transmitted to both output waveguides approximately equally and then detected by the SSPDs in the near-field. From these measurements we can conclude that the coupling ratio is 0.5 (see Supplementary Information).

To create pairs of indistinguishable plasmons we build a bright photon pair source based on a collinear type-II phase-matched degenerate Spontaneous Parametric Down-Conversion[25] process, depicted in Fig. 2b. The photon pairs ($\lambda$ = 1064 nm, Fig. S1) are separated at a polarizing beam-splitter and, after a motorized delay line in one of the arms, a single spatial mode is collected in single-mode fibers. We first characterize the indistinguishability of the produced photon pairs in a HOM experiment using a fiber beam splitter by performing a time-resolved correlation measurement. The recorded counts versus time, $t$, correspond to events where detector 1 clicks followed by a click from detector 2 a time $t$ later. For time-delays between the photons, $dt$, larger than 3 ps, the peak above the uncorrelated background in Fig. 4a around $t = 0$ shows that the counts come from a correlated source, i.e. there is an enhanced probability of simultaneous photon detection by the two detectors because our source emits photon pairs. When overlapping the photons in time at the fiber beam splitter by precisely adjusting the time-delay $dt$, the correlation peak disappears. In this case the photons bunch, i.e. interfere and travel to the same detector, such that there are no correlated clicks between the detectors. In Fig. 4b we plot the integrated area under the curves in Fig. 4a, resulting in the characteristic HOM dip. From the width and depth of the HOM dip we obtain a coherence time $\tau_c = (2.51 \pm 0.03)$ ps and an interference visibility $V = 0.92 \pm 0.01$, limited by non-perfect polarization and spectral overlap.

Next, the H- and V-polarized photons from the photon pair source are individually coupled to the low-temperature setup where they are focused to small separate spots (spot-size ~1 μm). After precisely aligning one spot at the left input waveguide and the other at the right waveguide (separation ~6 μm) and setting the half-wave plates for optimal plasmon excitation, the count-rates of the SSPDs are approximately 45 kHz. The time-resolved correlation data in Fig. 4c now shows a much narrower and more pronounced peak above the flat background, again because the detection events come from a correlated source. The improved signal to noise ratio is due to the increased time-resolution of the SSPDs, which effectively reduces the background signal intensity. We now adjust the time-delay between the two plasmons, and again see a decrease in correlated events. The integrated data in Fig. 4d shows the plasmonic HOM dip measured on two different devices, with $\tau_c$ = (2.60 ± 0.11) ps, $V$ = 0.43 ± 0.02 for device 1 and $\tau_c$ = (2.49 ± 0.08) ps, $V$ = 0.39 ± 0.01 for device 2. The interference visibility is reduced by about a factor 2 in both of the devices we measured. This reduction is due to the simultaneous excitation of long-range surface plasmon modes[26] which are not strongly confined and do not have cut-off. From a polarization-dependent plasmon excitation measurement (Fig. S2) we estimate the long-range surface plasmon mode population to be 0.19, which directly translates into a reduced quantum interference visibility $V$ = 0.49. Calculations show that imperfect mode overlap in the beam splitter reduces this value further by 8% (see Supplementary Information), bringing the expected value to 0.45, very close to the measured visibilities. However, it is important to observe that the width of the dip is identical to the coherence time of the photon pair source, implying that the temporal properties of the wave-packet are not affected by propagation in the plasmonic mode.

The plasmonic directional couplers we designed in this experiment are significantly smaller than previously realized structures[21,22,23] and allow further development of sub-wavelength integrated optical networks with built-in light-sources and detectors. This platform provides opportunities to efficiently extract and manipulate photons from single-photon emitters on-chip, without ever converting to a free-space photon. Most importantly, the observation of two-plasmon quantum interference proves that plasmon indistinguishability is maintained and therefore opens the way to implement quantum optical elements at the length scale of typical electronic components.

## Methods

*Simulations and device design*

We performed numerical simulations using an optical mode solver[27]. Mode-dispersion calculations for a single 150 nm thick waveguide (Fig. S5) show that below a width of ~500 nm this structure has a single strongly confined mode and two weakly confined long-range modes. We choose a 300 nm wide, 150 nm thick waveguide as the basic element of our plasmonic circuit (mode profile in Fig. S6).

In directional couplers, the interaction of the evanescent fields of two nearby waveguides causes their eigenmodes to hybridize into the symmetric ($|S\rangle$, Fig. S7) and anti-symmetric ($|AS\rangle$, Fig. S8) combinations called supermodes. Incoming single modes $|L\rangle$ and $|R\rangle$ in the left and right waveguides can be written in the supermode basis: $|L\rangle = 1/\sqrt{2}\ (|S\rangle + |AS\rangle)$, $|R\rangle = 1/\sqrt{2}\ (|S\rangle - |AS\rangle)$. Due to the effective index difference $\Delta n$ between modes $|A\rangle$ and $|AS\rangle$, the state after propagating over a length $\Delta x$ is given by $|\psi\rangle = |S\rangle + e^{i2\pi\Delta n\Delta x/\lambda}|AS\rangle$, with $\lambda$ the free-space wavelength. When separating this state into modes $|L\rangle$ and $|R\rangle$, the output power oscillates sinusoidally between the left and right waveguide as a function of $\Delta x$ (measurements in Fig. S3). In the case of plasmonic structures, losses give an overall damping factor (calculations in Fig. S10). Based on a trade-off between losses and $\Delta n$ we selected 250 nm wide waveguides with a 100 nm spacing to show directional coupler behavior. The device for the interference measurements had a gap of 150 nm and $L = 0$ to operate in the first 50 / 50 point.

*Detection efficiency*

The pair production-rate of the downconversion source was estimated to be 288 MHz with a fiber collection efficiency of ~35%, based on the measured single- and pair-rates[28] using silicon avalanche photodiodes with a detection efficiency of ~1.5% at 1064 nm. Taking into account the transmittance of the objective of ~55% at this wavelength, the overall efficiency of exciting and detecting a plasmon was ~8 x $10^{-4}$. From simulations and rough calculations we estimate that scattering into the plasmonic mode by focusing a laser at the end of a waveguide has an efficiency of order 5 - 10%.

**Author contributions**

R.W.H. designed the experiment, fabricated the samples and performed the measurements and the analysis. L.P.K. and V.Z. supervised the project. All authors contributed to the manuscript.

**References**

1. Barnes, W., Dereux, A. & Ebbesen, T. Surface plasmon subwavelength optics, *Nature* **424,** 824-830 (2003).
2. Schuller, J. *et al.* Plasmonics for extreme light concentration and manipulation, *Nature Mater.* **9**, 193-204 (2010).
3. Chang, D.E., Sørensen, A.S., Hemmer, P.R. & Lukin, M.D. Quantum optics with surface plasmons, *Phys. Rev. Lett.* **97**, 53002 (2006).
4. Oulton, R. *et al.* Plasmon lasers at deep subwavelength scale, *Nature* **461**, 629-632 (2009).
5. Schietinger, S., Barth, M., Aichele, T. & Benson, O. Plasmon-Enhanced Single Photon Emission from a Nanoassembled Metal-Diamond Hybrid Structure at Room Temperature, *Nano Lett.* **9**, 1694-1698 (2009)
6. Curto, A.G. *et al.* Unidirectional emission of a quantum dot coupled to a nanoantenna, *Science* **329**, 930-933 (2010).
7. Neutens, P., Van Dorpe, P., De Vlaminck, I., Lagae, L. & Borghs, G. Electrical detection of confined gap plasmons in metal-insulator-metal waveguides, *Nature Photon.* **3**, 283-286 (2009).
8. Falk, A. *et al.* Near-field electrical detection of optical plasmons and single-plasmon sources, *Nature Phys.* **5**, 475-479 (2009).
9. Akimov, A. *et al.* Generation of single optical plasmons in metallic nanowires coupled to quantum dots, *Nature* **450**, 402-406 (2007).
10. Kolesov, R. *et al.* Wave particle duality of single surface plasmon polaritons, *Nature Phys.* **5**, 470-474 (2009).
11. Altewischer, E., Van Exter, M. & Woerdman, J. Plasmon-assisted transmission of entangled photons, *Nature* **418**, 304-306 (2002).
12. Fasel, S. *et al.* Energy-time entanglement preservation in plasmon-assisted light transmission, *Phys. Rev. Lett.* **94**, 110501 (2005).
13. Fujii, G. *et al.*, Preservation of photon indistinguishability after transmission through surface-plasmon-polariton waveguide. *Opt. Lett.* **37**, 1535-1537 (2012).
14. Chang, D., Sørensen, A., Demler, E. & Lukin, M. A single-photon transistor using nanoscale surface plasmons, *Nature Phys.* **3**, 807-812 (2007).
15. Gol'tsman, G. *et al.* Picosecond superconducting single-photon optical detector, *Appl. Phys. Lett.* **79**, 705 (2001).
16. Heeres, R.W. *et al.* On-chip single plasmon detection, *Nano Lett.* **10**, 661-664 (2010).
17. Hong, C., Ou, Z. & Mandel, L. Measurement of subpicosecond time intervals between two photons by interference, *Phys. Rev. Lett.* **59**, 2044-2046 (1987).
18. Knill, E., Laflamme, R. & Milburn, G. A scheme for efficient quantum computation with linear optics, *Nature* **409**, 46-52 (2001).
19. Loudon, R. Fermion and boson beam-splitter statistics. *Phys. Rev. A* **58**, 4904 (1998).
20. Barnett, S.M., Jeffers, J., Gatti, A. & Loudon, R. Quantum optics of lossy beam splitters, *Phys. Rev. A* **57**, 2134 (1998).
21. Yamada, H., Chu, T., Ishida, S. & Arakawa, Y. Optical directional coupler based on Si-wire waveguides, *IEEE Photon. Technol. Lett.* **17**, 585-587 (2005).
22. Politi, A., Cryan, M. J., Rarity J. G., Yu, S. & O'Brien, J. L. Silica-on-Silicon Waveguide Quantum Circuits, *Science* **320**, 646-649 (2008).
23. Charbonneau, R., Lahoud, N., Mattiussi, G. & Berini, P. Demonstration of integrated optics elements based on long-ranging surface plasmon polaritons, *Opt. Express* **13**, 977-984 (2005).
24. Bozhevolnyi, S., Volkov, V., Devaux, E., Laluet, J. & Ebbesen, T. Channel plasmon subwavelength waveguide components including interferometers and ring resonators, *Nature* **440**, 508-511 (2006).
25. Burnham, D. & Weinberg, D. Observation of simultaneity in parametric production of optical photon pairs, *Phys. Rev. Lett.* **25**, 84-87 (1970).
26. Jung, J., & Søndergaard, T. & Bozhevolnyi, S.I. Theoretical analysis of square surface plasmon-polariton waveguides for long-range polarization-independent waveguiding. *Phys. Rev. B* **76**, 035434 (2007).
27. Lusse, P., Stuwe, P., Schule, J. & Unger, H. Analysis of vectorial mode fields in optical waveguides by a new finite difference method, *J. Lightwave Technol.* **12**, 487-494 (1994).
28. Hadfield, R. Single-photon detectors for optical quantum information applications, *Nature Photon.* **3**, 696-705 (2009).

**Figure captions**

**Figure 1 | Hong-Ou-Mandel interference. a**, Conventional two-photon quantum interference. Pairs of photons are created in a parametric downconversion process and collected in two optical fibers. Quantum interference takes place in a fiber-based beam splitter. Photons are measured using silicon APDs. **b**, On-chip plasmon interference. Photon pairs are generated as in **a**, the interference and detection part of the setup is several orders of magnitude smaller. A plasmonic directional coupler functions as beam splitter. SSPDs are integrated to perform near-field detection. The dead time and jitter of SSPDs is much smaller than that of usual APDs, as indicated schematically by insets of the output signals.

**Figure 2 | Plasmon interference device and experimental setup. a**, SEM image of a gold plasmonic directional coupler device. Integrated SSPDs (colourised in yellow) are located ~20 nm below the output waveguides. Top-right: zoom-in of the plasmon detection region. The width and spacing of the superconducting meander are both 100 nm, thickness ~5 nm. The 150 nm thick gold waveguide is 600 nm wide and aligned for optimal absorption of the plasmonic mode in the SSPD (Fig. S11). Bottom-right: zoom-in of the coupling region, waveguide width is 250 nm, gap between waveguides is 100 nm. The cross coupling ratio can be controlled by varying the interaction length $L$. **b**, Photon-pair source based on degenerate type-II collinear spontaneous parametric downconversion. A ~400 mW pump laser (532 nm) is focused to a ~70 μm diameter spot in a 2 cm long Potassium Titanyl Phosphate (KTP) crystal producing photon pairs. The polarizing beam splitter (PBS) separates the H and V-polarized photons from a pair to be collected in different optical fibers. A motorized delay line in one of the arms allows to control the time-delay between the two photons with ~16 fs resolution. **c**, Experimental setup. Two beams from optical fibers are focused to separate spots (spot size ~1 μm diameter) on the sample by a microscope objective. Their polarization can be rotated using half-wave plates to optimize plasmon excitation (Fig. S2). Objective and sample are at 4 K in a dipstick filled with Helium exchange gas and with a window for free-space optical access.

**Figure 3 | Photo-response maps of a plasmonic beam splitter. a**, Measurement procedure: a focused laser beam is raster-scanned across the device. At each location we wait for 0.2 seconds and count the number of clicks in both the left and right detector. **b, c**, Photo-response maps of a 50/50 plasmonic directional coupler with two SSPDs (device as in Fig. 2a, but with 150 nm gap and $L = 0$). The colour at each pixel represents the number of clicks obtained in the left (**b**) or the right (**c**) detector when the laser excitation is at the location of that pixel. The top-left of **b** and top-right of **c** show that each of the detectors individually click very often when directly illuminated, i.e. when absorbing photons. The two spots at the bottom of **b** and **c** correspond to plasmons being excited at the ends of the waveguides and reaching both detectors with almost equal intensity implying a 50/50 coupling strength. Note that a different color scale is used in the upper and lower parts in **b** and **c** since plasmon excitation yields less counts due to a limited photon-plasmon coupling efficiency and propagation losses in the waveguide. The laser is linearly polarized and set to excite plasmons with equal intensity at both inputs (approximately vertical with respect to this image). Structure design superimposed as a guide to the eye.

**Figure 4 | Quantum interference measurements. a**, Time-resolved correlation measurements to characterize the photon pair source using the scheme in Fig. 1a (fiber beam splitter-based). The x-axis spans the range -4 to 4 ns, much larger than the inter-photon delay, *dt*, indicated for each curve. For an inter-photon delay of more than 3 ps the central peak around $t = 0$, well above the uncorrelated flat background, corresponds to a second order correlation function $g^{(2)}(0)$ larger than 1 due to the correlated source. The peak width is limited by a jitter of ~600 ps in the APDs. Quantum interference causes these correlations to disappear almost completely when the photons overlap in time. Integration time per curve is 8 seconds. **b**, HOM interference dip of the photon pair source. The coincidences on the vertical axis are equal to the integrated number of counts at particular delay times, *dt*, between the two photons from a single pair (i.e. shaded areas in **a**). The full width at half maximum of the Gaussian fit to the HOM-dip gives a coherence time $\tau_c = (2.51 \pm 0.03)$ ps and visibility $V = 0.92 \pm 0.01$. **c**, Time-resolved correlation measurements of plasmon interference using the scheme in Fig. 1b. The x-axis now only spans from -1 to 1 ns because the SSPDs have significantly improved timing resolution (jitter ~120 ps). This results in a lower background and a correspondingly larger $g^{(2)}(0)$. (The time resolution is still larger than the indicated inter-plasmon delay). Integration time 15 minutes per curve. **d**, HOM interference dip showing quantum interference of surface plasmons in two devices (as shown in Fig. 2a) with a 50/50 directional coupling strength (gap 150 nm, $L = 0$). Coherence time $\tau_c = (2.60 \pm 0.11)$ ps, visibility $V = 0.43 \pm 0.02$ for device 1 (integration time 30 min/point, pairs of traces from **c** combined) and $\tau_c = (2.49 \pm 0.08)$ ps, $V = 0.39 \pm 0.01$ for device 2 (integration time 20 min/point). Compared to the photon pair source properties in **b** the measured coherence time is identical but the visibility is reduced.

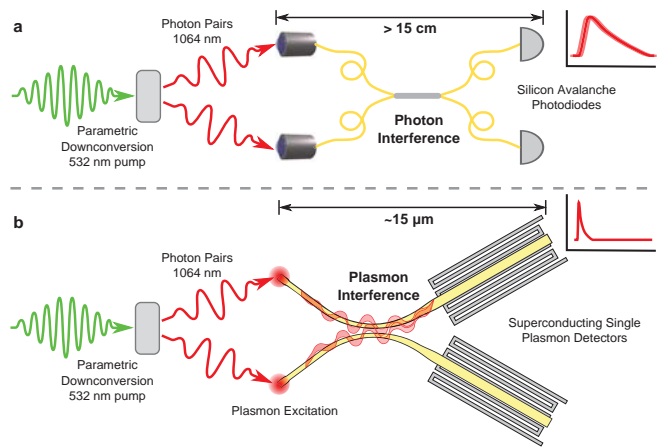

Figure 1

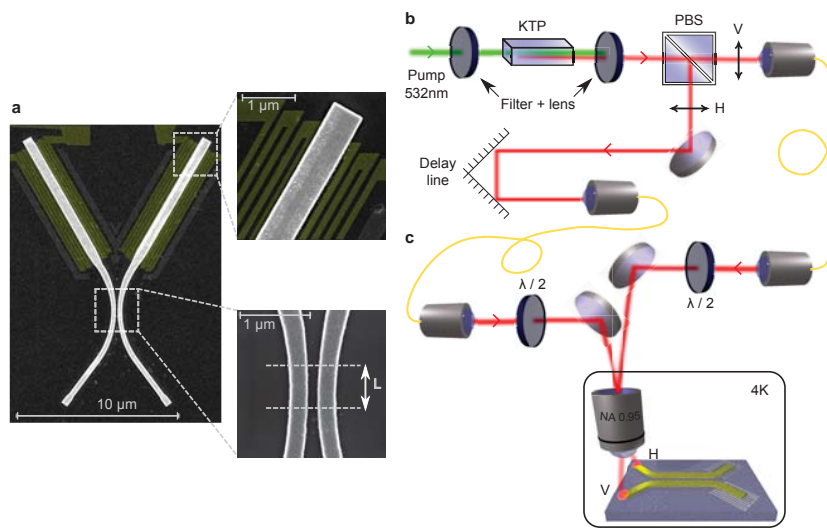

Figure 2

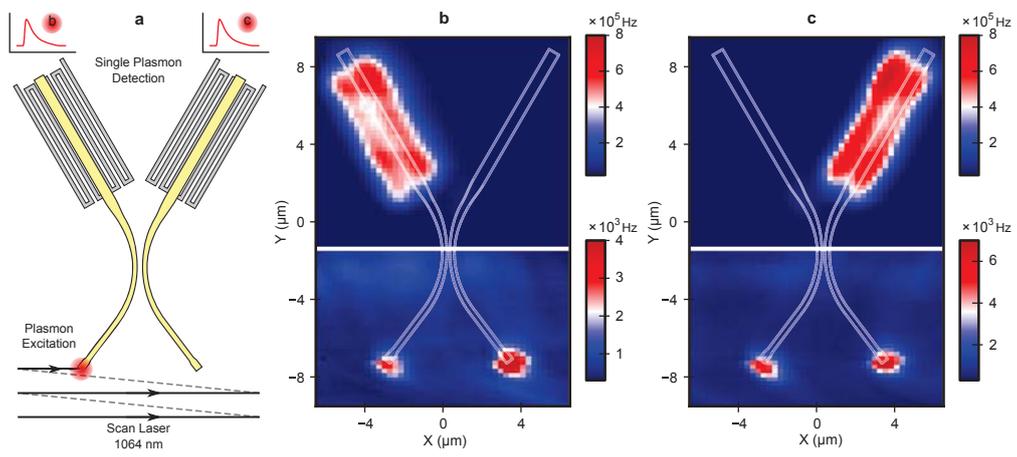

Figure 3

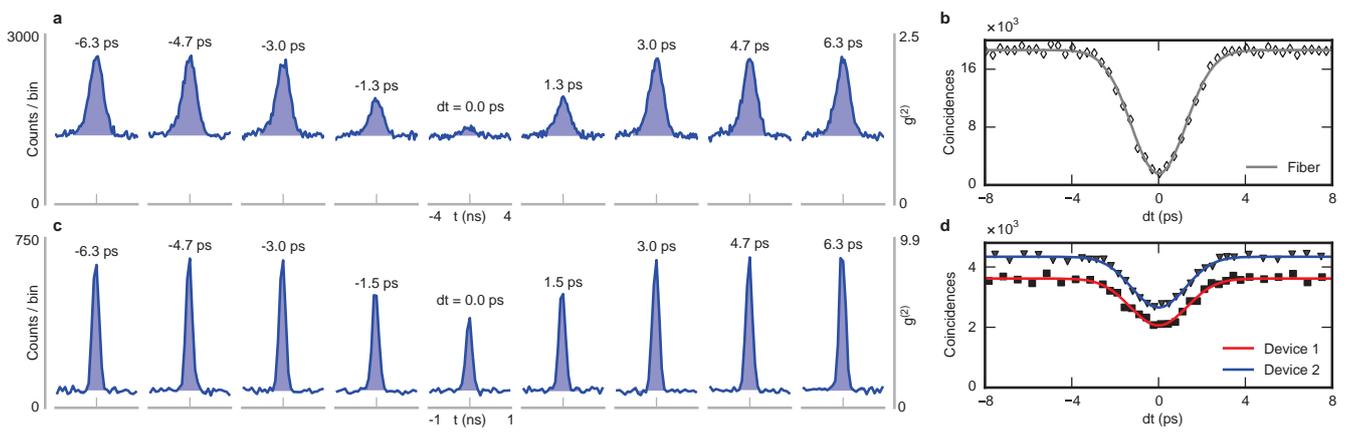

Figure 4


\newpage

\tableofcontents
\appendix
\newpage

\section{Experimental setup}

\subsection{Spontaneous parametric downconversion source}
\begin{figure}[bh]
\begin{center}
        \mbox{
		\includegraphics{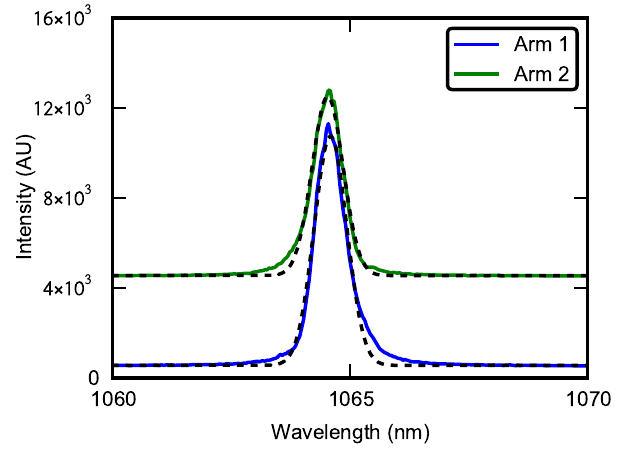}
	}
\end{center}
\caption{\label{fig:downconversion} \textbf{Downconversion spectra.}
Spectra collected in output arm 1 and 2 of the downconversion setup with calculated overlap of $98.7\%$. Gaussian fits have a FWHM of $(809 \pm 3) \unit{pm}$ and $(770 \pm 3) \unit{pm}$ respectively. Taking $\Delta f = \frac{c}{\lambda^2}\Delta \lambda$ the expected coherence time $\tau_c \approx \frac{1}{\Delta f} \approx 4.8 \unit{ps}$.}
\end{figure}

The photon pair source is based on a $2 \unit{cm}$ long Potassium Titanyl Phosphate (KTP) crystal, phase-matched for degenerate collinear type-II downconversion from $532 \unit{nm}$ to $1064 \unit{nm}$ and pumped with $\sim 400 \unit{mW}$ of $532 \unit{nm}$ laser light from a Millenia diode pumped solid state laser (beam diameter $2.3 \unit{mm}$). This beam still has some background radiation at $1064 \unit{nm}$, which is short-pass filtered (Semrock FF01-529/24-25 and Thorlabs FL532-10). After the crystal, the pump light is filtered out using a long pass filter (Semrock BLP01-635R-25). The spectra in both arms after the polarizing beam splitter are shown in \reffig{fig:downconversion}. The spectral width is set by the length of the crystal in combination with the phase-matching condition. Care is taken to use the optimal focusing condition \cite{fedrizzi2007wavelength} by choosing the lenses such that the crystal length is about 2 times the Rayleigh range for the pump beam, and 3.5 times the Rayleigh range for the signal and idler beam. In our case we use a focusing lens with $f = 250 \unit{mm}$ and a collection lens with $f = 175 \unit{mm}$.

Correlation measurements on APDs (Perkin Elmer SPCM-AQRH) with an efficiency of $\sim 1.5\%$ gave single count rates of $1.45 \unit{MHz}$ and $1.6 \unit{MHz}$ and a pair rate of $8050 \unit{Hz}$, giving an estimated pair production-rate of $\frac{N_1N_2}{N_c} = 288 \unit{MHz}$.

\subsection{Sample fabrication}
The samples are fabricated on a sapphire substrate purchased from Scontel with $\sim 5 \unit{nm}$ NbN sputtered with the substrate heated to $\sim 800$\textcelsius. Contacts are defined by e-beam writing in a $\sim 250 \unit{nm}$ thick PMMA layer followed by evaporation and lift-off of chrome ($15 \unit{nm}$) / gold ($50 \unit{nm}$). The SSPDs are patterned by e-beam in a $70 \unit{nm}$ thick hydrogen silsesquioxane (HSQ) layer, developed in TMAH ($5 \unit{sec}$) and $\mathrm{MF322:H_2O}$ 1:9 ($15 \unit{sec}$) and used as an etching mask for an $\mathrm{SF_6}$ reactive ion etch (RIE). Usually the HSQ is removed by a $2 \unit{sec}$ dip in BHF, but for the quantum interference sample the remaining layer ($\sim 40 \unit{nm}$ after etching) was left on as it seemed to damage some devices, probably due to dirt under the NbN film. A thin insulating layer of $\sim 10 \unit{nm}$ $\mathrm{Al_2O_3}$ is deposited by Atomic Layer Deposition (ALD). The waveguides require a three layer mask because the substrate is not conductive at this point. Therefore we spin a $\sim 450 \unit{nm}$ layer of Shipley S1805 photo-resist, sputter a $\sim 10 \unit{nm}$ layer of tungsten and spin a layer $\sim 90 \unit{nm}$ PMMA. The waveguide pattern is e-beam written in the PMMA. After developing in MIBK:IPA 1:3, the tungsten is removed with an $\mathrm{SF_6}$ RIE and the S1805 using an $\mathrm{O_2}$ plasma. This makes sure the substrate is very clean, so a $150 \unit{nm}$ gold layer can be evaporated without using a sticking layer, which would make the plasmonic modes much more lossy. Lift-off is performed in warm acetone and followed by an $\mathrm{HNO_3}$ dip ($5 \unit{sec}$) to remove resist residues. The waveguides are covered with $\sim 10 \unit{nm}$ ALD $\mathrm{Al_2O_3}$. The last step is to add a thick $\mathrm{Al_2O_3}$ layer to make the environment more symmetric. This is done by sputtering a $\sim 7 \unit{nm}$ layer of chrome and spinning a $\sim 900 \unit{nm}$ thick layer of PMMA. After e-beam writing and developing the chromium layer is removed by wet etching and a $\sim 550 \unit{nm}$ layer of $\mathrm{Al_2O_3}$ is sputtered. Finally, lift-off is performed in acetone and the remaining chromium is removed by wet etching.

\subsection{Measurement setup}
The SSPDs are biased using a home-made bias tee with built-in RC filters. The high frequency output is amplified by a $1 \unit{GHz}$ bandwidth minicircuits ZFL-1000LN+ followed by a $1.45 \unit{GHz}$ RF-Bay LNA-1450. For pulse counting measurements the signal is converted to TTL pulses using a comparator circuit and sent to a frequency divider. The divide-by-2 output is connected to a National Instruments USB-6216 card for counting. Correlations are measured by directly sending the generated TTL pulses to a Picoharp 300 time correlated single photon counter.

The sample is mounted on an Attocube XYZ slip-stick piezo positioner stage. For the measurements in Fig. 3 of the main text, actually the sample is raster scanned and the focused laser spot is fixed.

\begin{figure}[ptbh]
\begin{center}
	\mbox{
		\includegraphics{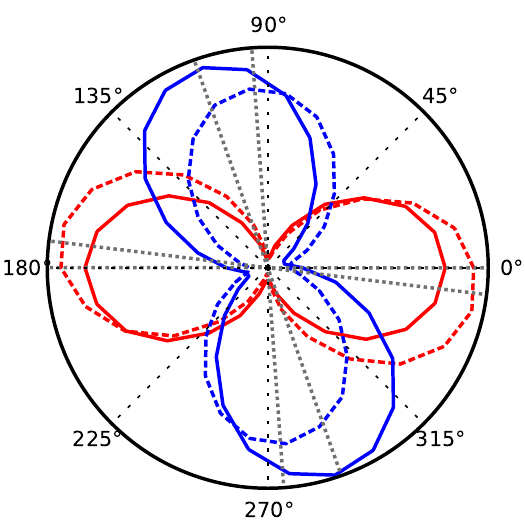}
	}
\end{center}
\caption{\label{fig:poldep} \textbf{Polarization dependence.}
Polarization dependence of plasmon excitation. The colors red and blue correspond to illumination of the left and right input respectively. Solid and dashed lines represent signal intensity of the left and right detector. The dotted gray lines indicate the angles that gives most efficient excitation and are obtained from a fit.
}
\end{figure}

\section{Additional measurements}

\subsection{Polarization dependence}
\label{sec:poldep}
To characterize the plasmon excitation efficiency we have measured the dependence of the detector signals as a function of the polarization angle of a linearly polarized laser beam illuminating one input at a time. The result of this measurement is shown in \reffig{fig:poldep}. There is a clear optimum excitation angle, different for both of the inputs which are designed to be oriented at a 75\textdegree\ angle with respect to each other. However, both the fact that the visibility is not 100\% and the fact that the two detectors do not show an identical polarization dependence indicates that we are exciting multiple modes. We are mostly interested in the strongly confined mode, but the long-range surface plasmon modes (LRSP, see section \ref{sec:simulations}) are also excited. We can estimate the population of the LRSP modes from the difference in the optimum polarization angle, which is on average 11\textdegree \ (in 2 devices that we have measured). The population of other modes then corresponds to a fraction of $\sin{11} \approx 0.19$.

\subsection{Cross coupling factor}
\begin{figure}[phtb]
\begin{center}
	\begin{tabular}{c c}
		\mbox{
			\includegraphics{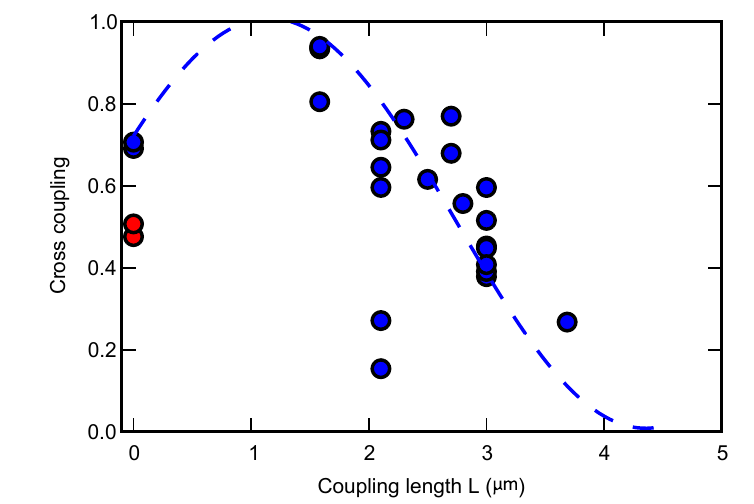}
		} &
		\mbox{
			\includegraphics[width=6cm]{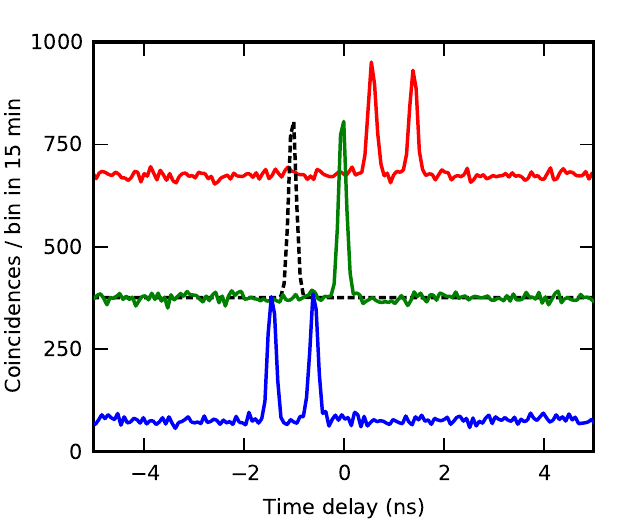}
		} \\
		\bf{(a)} & \bf{(b)}
	\end{tabular}
\end{center}
\caption{\label{fig:splitratio}
(a) Measured cross coupling factors for different directional couplers as a function of coupling length L. Blue dots correspond to devices with a gap of $100 \unit{nm}$, red dots to devices with a gap of $150 \unit{nm}$. The dashed line is a guide to the eye indicating a typical interaction length $L_{\sfrac{\pi}{2}} = 1.6 \unit{\mu m}$.
(b) Time-resolved correlation measurements (bin size $64 \unit{ps}$) between detection events of two SSPDs in a directional coupler device (gap $150 \unit{nm}$, $L = 0$, one of the red dots in (a)). Large time delay between the 2 correlated plasmons (blue and red curve) and small time delay (green curve). The dashed black curve is a Gaussian fit with a FWHM of $173 \unit{ps}$ implying a jitter of $173 / \sqrt{2} = 122 \unit{ps}$. Curves offset both in x and y for clarity.
}
\end{figure}

The cross coupling factor $c$ is estimated by successively illuminating the left and right input with a focused laser spot and optimizing the detector count-rates at both positions. Considering $c$ to be identical for left-to-right and right-to-left coupling, the count rates are given by $C_l = I_l(1 - c)\eta_l$ and $C_r = I_lc\eta_r$ when exciting plasmons in arm $l$ with intensity $I_l$. When exciting the right arm the count rates are given by $C'_l = I_rc\eta_l$ and $C'_r = I_r(1 - c)\eta_r$. From these equations the coupling factor can be obtained, as well as the relative detector and plasmon excitation efficiencies. We have measured the coupling factor on many devices with different coupling length $L$ and show the results in \reffig{fig:splitratio}a. The characteristic length for a 50/50 splitter based on two waveguides with a $100 \unit{nm}$ gap is estimated to be $L_{\sfrac{\pi}{2}} \approx 1.6 \unit{\mu m}$ from the dashed line. Important to note is that a splitter can be realized in which almost all of the power is coupled from one arm to the opposite arm, confirming that these devices function as directional couplers. The dashed curve indicates that this should happen for $L \approx 1.2 \unit{\mu m}$. Although we did not fabricate devices of exactly that length, the measurements with $L \approx 1.6 \unit{\mu m}$ show a cross coupling close to one. The devices with $L = 0$ have no extra coupling length, so only consist of the waveguides smoothly coming together until the gap size and separating again. This already results in a significant coupling between them. They seem to show more uniform cross coupling behavior when comparing to the other, longer devices. The spread in cross coupling for these longer devices is caused by the fact that it is hard to create a uniform gap of $100 \unit{nm}$ between two waveguides over a length of micrometers. The irregularities, for example due to the grain structure and other fabrication imperfections, will result in a spread of cross coupling factors.

A more accurate measurement of the cross coupling factor can be performed using the photon pair source. We set a large delay between the photons forming a pair and send one of them to the left input and the other to the right input. By performing a time-resolved correlation measurement we can observe the correlated clicks of the left and right detector. These will only show correlated events when either both plasmons forming a pair stay in the arm where they were excited, or when both of them cross to the opposite arm. Because of the large time delay between the plasmons these events can be distinguished in time, as can be seen in the upper and lower trace in \reffig{fig:splitratio}b. The area of the peak due to plasmons coupling to the opposite arm scales with $c^2$, the other one with $(1 - c)^2$. The fact that the peaks are almost equal in area confirms that $c$ is close to $0.5$ and therefore that this is a 50/50 coupler.

\begin{figure}[phtb]
\begin{center}
	\mbox{
		\includegraphics{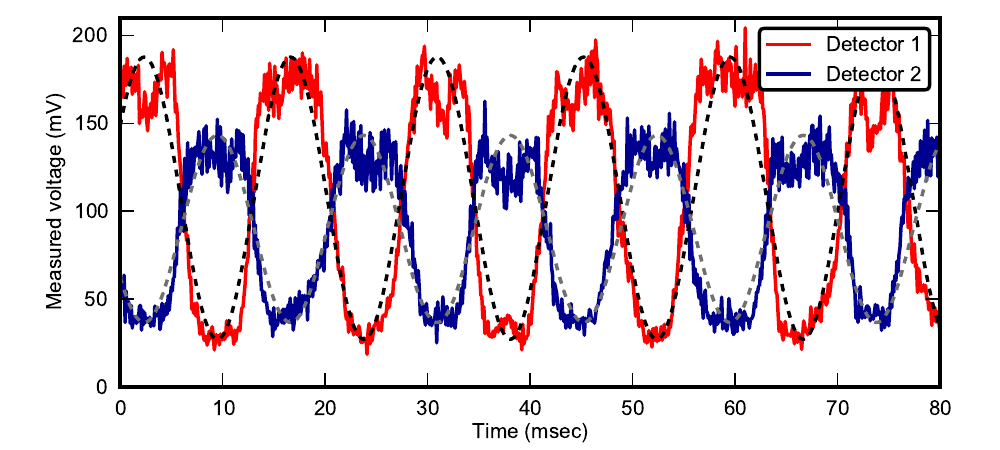}
	}
\end{center}
\caption{\label{fig:classical} \textbf{Classical interference. }
Analog output signal of the detectors at the beam splitter outputs (see text) as a function of time. Both input waveguides are excited using a 1064 nm laser beam with an oscillating phase between them ($70 \unit{Hz}$), applied using a mirror on a piezo-electric stage. The interference signal at detector 1 and 2 has a visibility of 0.747 and 0.592 respectively.
}
\end{figure}

\section{Classical interference}
To perform a classical interference experiment we first convert the SSPD signal pulses to an analog signal. A comparator is used to amplify and stretch each detection pulse to a 5V, $\sim 50 \unit{ns}$ long signal. A low pass filter with a cut-off frequency of about $5 \unit{kHz}$ is used to form the output. The two input waveguides are then simultaneously excited using two separate laser spots, originating from the same laser but with a periodic phase modulation (70 Hz) applied to one of the beams using a mirror on a piezo-electric stage. The resulting detector signals are shown in \reffig{fig:classical} and clearly show a classical interference signal, although with reduced visibility $V = 0.747$ for detector 1 and $V = 0.592$ for detector 2. The reduced and unequal visibility in the signals is due to the excitation of the LRSP modes and imperfect mode overlap in the coupling area. For the interference visibility in the quantum experiment we expect an upper bound given by the average of the classical visibilities, or $V = 0.669$. It should be noted that the two detector signals are $\pi$ out of phase, which means that the ideal beam splitter phase relations are properly maintained.

\section{Quantum interference visibility}
Our limited quantum interference visibility can be explained by the population of the LRSP modes that our waveguides supports (section \ref{sec:simulations}). In section \ref{sec:poldep} we estimated the population of these 2 modes to be $P \approx 0.19$. The input state of our beam splitter can now be described as:
\begin{eqnarray}
 \ket{\psi} &=& (\sqrt{1 - P} \ket{S_1} + \sqrt{P} \ket{L_1}) \times (\sqrt{1 - P} \ket{S_2} + \sqrt{P} \ket{L_2}) \\
           &=& (1 - P) \ket{S_1S_2} + P \ket{L_1L_2} + \sqrt{P(1 - P)}(\ket{L_1S_2} + \ket{S_1L_2})
\end{eqnarray}
where $S_i$ and $L_i$ respectively indicate the strongly confined and LRSP mode in input $i$. Now only the first term will result in quantum interference, with an intensity $I = (1 - P)^2$. Against a background of $1 - I$ this directly results in a visibility $V = \sfrac{(max - min)}{(max + min)} = \sfrac{I}{(2 - I)} \approx 0.49$. The imperfect overlap of the single waveguide mode with the supermodes in the coupling region (see section \ref{sec:simulations}) will degrade the interference visibility further by 8\%, bringing the total expected visibility $V$ down to $0.45$, very close to the value we observe in our experiment.

A Hong-Ou-Mandel like interference dip with a visibility of up to 0.5 can be obtained for classical fields if they are coherent. This HOM-\textit{like} effect can be used to measure the visibility of the first order correlation function \cite{ou1989fourth}. In the present experiment, however, the two beams from the downconversion process are mutually incoherent because we operate in the spontaneous, or low-photon number, regime \cite{joobeur1994spatiotemporal}. Therefore the HOM-\textit{like} effect can not account for our observations. From the time-resolved correlation measurements we also conclude that the bunching only occurs at very short time-scales, whereas the HOM-\textit{like} effect would result in bunching on the time-scale of the phase-fluctuations. By showing that we can observe classical interference at $70 \unit{Hz}$ we prove that the phase-fluctuations in our setup are much slower than the bunching time-scale of less than a nanosecond that we observe (still limited by our detectors).

Another clear aspect of the quantum mechanical nature of the plasmons in this experiment can be found in the correlation statistics: the two beams combined yield statistics very different from the product of the statistics of the individual beams. Whereas each of the beams obeys Poissonian photon statistics, the cross correlation shows very strong bunching. This is essentially a manifestation of photon-number entanglement between the two beams and no classical field can mimic this property \cite{mandel1986non}. The heralded states we use make sure that if a photon is detected in one of the beams a partner photon is present in the other, which is therefore projected in the Fock-state $|1\rangle$, one of the clearest non-classical states.

\begin{figure}[bh]
\begin{center}
	\begin{tabular}{c c}
		\mbox{
		  \includegraphics{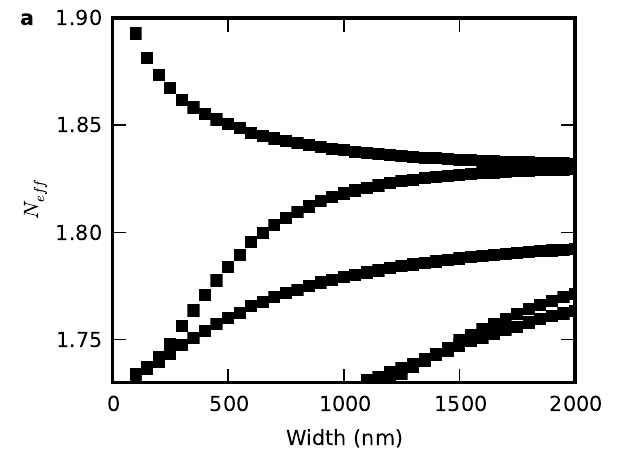}
		} &
		\mbox{
		  \includegraphics{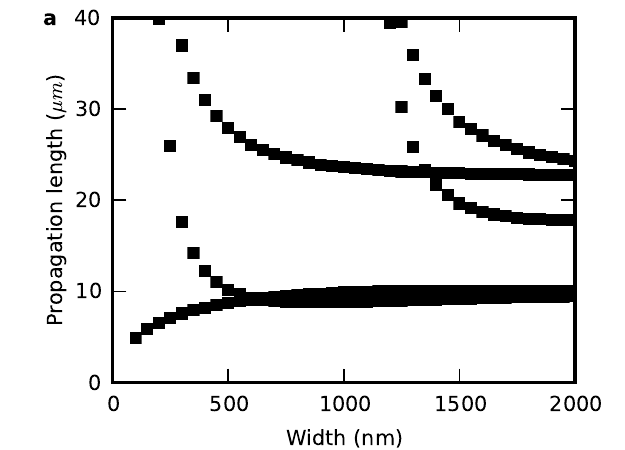}
		} \\
		\bf{(a)} & \bf{(b)}
	\end{tabular}
\end{center}
\caption{\label{fig:dispersion} \textbf{Mode dispersion. }
Effective index (a) and propagation length (b) as a function of waveguide width, for a gold waveguide of $150 \unit{nm}$ thick.
}
\end{figure}

\begin{figure}[tbph]
\simmode{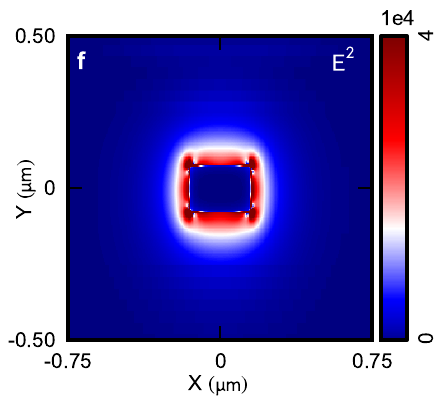}
\caption{\label{fig:single_mode} \textbf{Single waveguide mode. }
Eigenmode of a single gold waveguide (width $300 \unit{nm}$, thickness $150 \unit{nm}$) embedded in sapphire. Effective index $n_{eff} = \neffsingle$. Normalization: $max(H_x,H_y) = 1$}
\end{figure}

\section{Simulations}
\label{sec:simulations}

\subsection{Single waveguide}
An optical mode solver is used to find the eigenmodes of a single gold waveguide ($\epsilon = -52.0595 + 3.385i$, thickness $150 \unit{nm}$) on top of a sapphire substrate ($\epsilon = 3.06$). The structure is covered with a $500 \unit{nm}$ dielectric layer of the same index on top; above that is air. The dispersion relation and propagation lengths ($\delta = \frac{\lambda}{4 \pi Im(n)}$) are shown in \reffig{fig:dispersion}a and b respectively. Below a width of $\sim 1 \unit{\mu m}$ there is only one strongly confined mode, with a high effective index and correspondingly a relatively short propagation length. This mode will be referred to as the Short Range Surface Plasmon (SRSP) mode. The last two remaining modes with a decreasing effective index for narrower waveguides are long-range surface plasmon (LRSP) modes which are not well-confined \cite{jung2007theoretical}.

As input waveguides for our beam splitter structure we selected a $300 \unit{nm}$ wide waveguide. The different components of the SRSP mode are shown in \reffig{fig:single_mode}. Note that the mode solver code gives the magnetic field components of the eigenmodes and the electric field components are derived from those. This requires taking derivatives on the discrete grid and the resulting electric fields are not as smooth as the magnetic fields, i.e. contain some large-valued pixels which causes the scaling to look somewhat odd.

A problem of small waveguides is that they result in high losses, in this case a propagation length of $\sim 7.5 \unit{\mu m}$. After the beam we increase the width to 600 nm to give a slightly increased propagation length of $\sim 9 \unit{\mu m}$. This ensures that a larger fraction of the power is absorbed by the SSPD instead of the gold waveguide.

\begin{figure}[ptbh]
\begin{center}
	\mbox{
	  \includegraphics{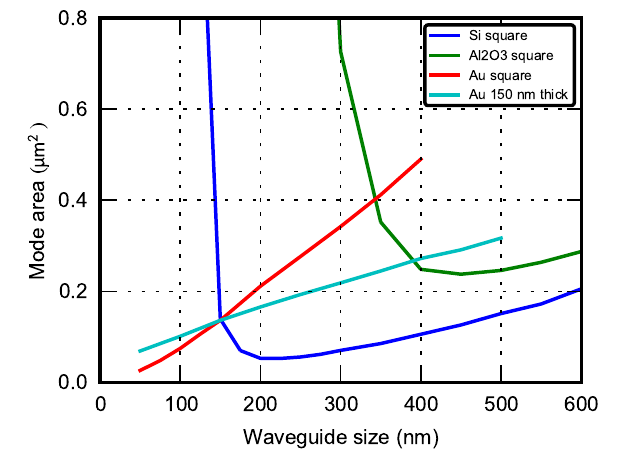}
	}
\end{center}
\caption{\label{fig:mode_area} \textbf{Effective mode areas.} Mode areas at $\lambda = 1064 \unit{nm}$ as a function of waveguide size calculated for plasmonic and dielectric waveguides.
}
\end{figure}

\subsection{Mode areas}
The waveguides we use offer confinement just below the diffraction limit of sapphire waveguides. To justify this statement we have calculated the effective mode area $\mathrm{A_{eff}}$, as is commonly used in fiber optics \cite{agrawal2001fiber}:
\begin{equation}
 \mathrm{A_{eff}} = \frac{ \left( \int |H|^2 dr \right)^2}{\int |H|^4 dr}
\end{equation}
We plot the effective area for a square gold waveguide and a gold waveguide of 150 nm thick in Fig. \ref{fig:mode_area}. We compare this to a dielectric waveguide made out of silicon (n = 3.6) embedded in $\mathrm{SiO_2}$ and to another one made of $\mathrm{Al_2O_3}$ (n = 1.75) embedded in air. We see that our structure results in a mode area that is just below the smallest possible mode in $\mathrm{Al_2O_3}$. The plasmonic structure size required to reach a mode area comparable to the diffraction limit of an $\mathrm{Al_2O_3}$ waveguide is about 2 times as small, as a result of the plasmonic mode being confined to the surface and not within the dielectric. Silicon, with its much higher dielectric constant allows for still smaller modes. What is important to note, however, is that the plasmonic modes offer a way to decrease the mode area further below the diffraction limit by reducing the waveguide size, although this mode will be lossier as well. Our calculated mode areas present an upper limit, as the electric energy is slightly more confined than the magnetic energy. Because the electric fields are deduced and therefore less accurate quantities in our simulations, we chose to use the magnetic energy density only to estimate $\mathrm{A_{eff}}$.

\begin{figure}[ph]
\simmode{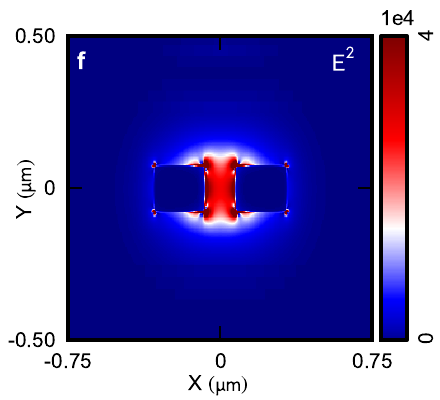}
\caption{\label{fig:coupled_symm} \textbf{Symmetric supermode. }
The symmetric (in $H_y$ and $E_x$) supermode profile of two coupled gold waveguides (width $250 \unit{nm}$, thickness $150 \unit{nm}$, gap $150 \unit{nm}$) embedded in sapphire. Effective index $n_{eff} = \neffsymmlarge$. For 100 nm gap: $n_{eff} = \neffsymmsmall$.
}

\simmode{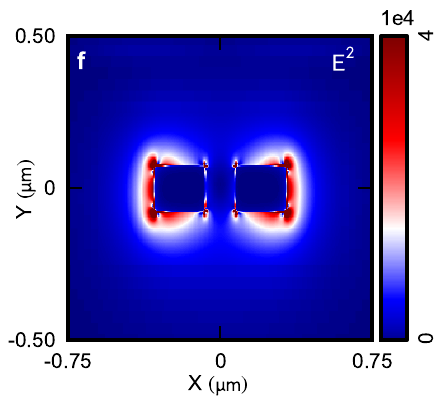}
\caption{\label{fig:coupled_antisymm} \textbf{Anti-symmetric supermode. }
The anti-symmetric (in $H_y$ and $E_x$) supermode profile of two coupled gold waveguides (width $250 \unit{nm}$, thickness $150 \unit{nm}$, gap $150 \unit{nm}$) embedded in sapphire.  Effective index $n_{eff} = \neffantisymmlarge$. For 100 nm gap: $n_{eff} = \neffantisymmsmall$.
}
\end{figure}

\subsection{Coupled waveguides}
\begin{figure}[tbph]
\begin{center}
	\begin{tabular}{c c}
                \mbox{
			\includegraphics{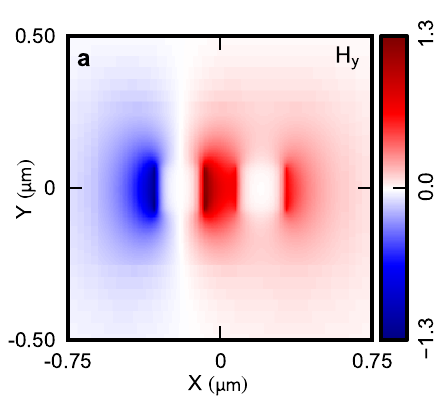}
		} & \mbox{
			\includegraphics{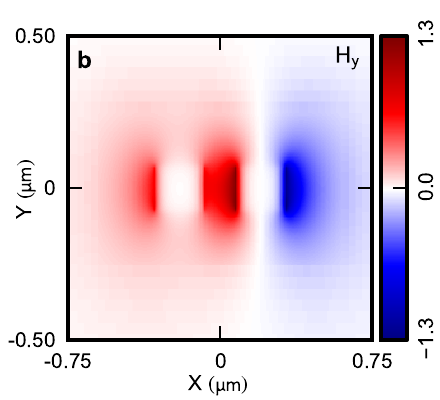}
		} \\
		\hspace{0.3cm}\bf{(a)} & \hspace{0.3cm}\bf{(b)}
	\end{tabular}
\end{center}
\caption{\label{fig:coupled_addsub} \textbf{Supermode combinations. }
Sum (a) and difference (b) of the supermodes in \reffig{fig:coupled_symm} and \reffig{fig:coupled_antisymm}. The results resemble the fundamental mode of a single waveguide in the left (a) and right (b) arm as in \reffig{fig:single_mode}, except for the extra field component in the opposite waveguide.
}
\end{figure}

The coupling region consists of $250 \unit{nm}$ wide, $150 \unit{nm}$ thick waveguides with a gap of 100 or 150 nm. The mode profiles of the symmetric and anti-symmetric modes for a structure with a 150 nm gap are shown in \reffig{fig:coupled_symm} and \reffig{fig:coupled_antisymm}. We have calculated overlap integrals with the single mode of a 250 nm wide waveguide, and find an overlap of $\mathrm{f_1} = 0.6962$ and $\mathrm{f_2 = \pm 0.6617}$ with the symmetric and anti-symmetric mode respectively. This leads to a fraction $\mathrm{f_1}^2 + \mathrm{f_2}^2 = 0.92$ of the energy being transmitted, and a fraction of $0.08$ being scattered. By increasing the distance between the two waveguides it is possible to reduce the amount of scattered light, although this will result in a smaller effective index difference and therefore a longer coupling section. In \reffig{fig:coupled_addsub}a and b we show the sum and difference of the symmetric and anti-symmetric mode. These images indeed resemble the single waveguide modes (\reffig{fig:single_mode}b) quite well, but show that the scattered light will mostly be on the far side, opposite of the waveguide that is being excited. This scattered light will lead to a decrease in interference visibility. The calculated 8\% scattered light is a lower bound: defects on our waveguides could increase this value. Although it is hard to estimate exactly how large this effect is, the scans presented in the main text show that only the ends of the waveguides are efficient plasmon in-couplers; we therefore do not expect too much out-coupling due to scattering either.

\begin{figure}[htbp]
\begin{center}
	\mbox{
		\includegraphics{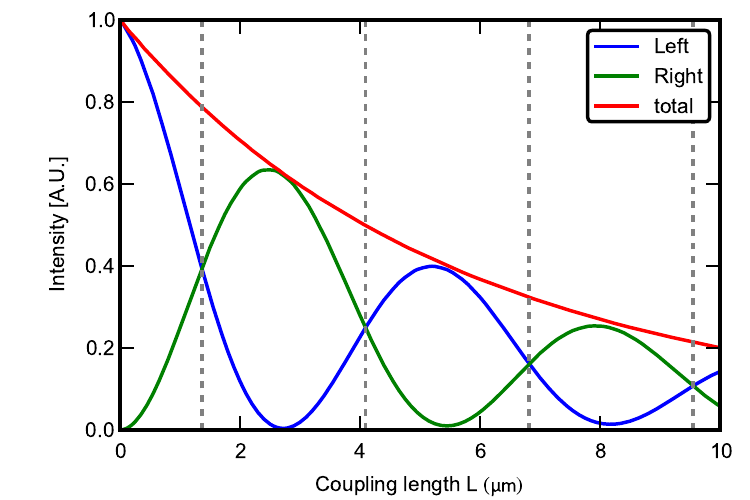}
	}
\end{center}
\caption{\label{fig:splitloss} \textbf{Beamsplitter loss. }
Output power in the left and the right arm of a directional coupler as a function of coupling length $L$, with effective indices of the modes in \reffig{fig:coupled_symm} and \reffig{fig:coupled_antisymm} with a 100 nm gap. The characteristic length $L_{\sfrac{\pi}{2}} = 1.36 \unit{\mu m}$.
}
\end{figure}

Using the complex effective indices of the supermodes we can calculate the expected cross-coupling and loss as a function of the directional coupler length; this is shown in \reffig{fig:splitloss}. This gives a characteristic length $L_{\sfrac{\pi}{2}}$ of $1.36 \unit{\mu m}$ for waveguides with a 100 nm gap, slightly less than the experimental value of $\sim 1.6 \unit{\mu m}$. The difference could be due to the gap size or waveguide width slightly deviating from the design or the real dielectric constants being different than the values used in the simulation.

\begin{figure}[thbp]
\begin{center}
	\begin{tabular}{c c c}
		\mbox{
		  \includegraphics{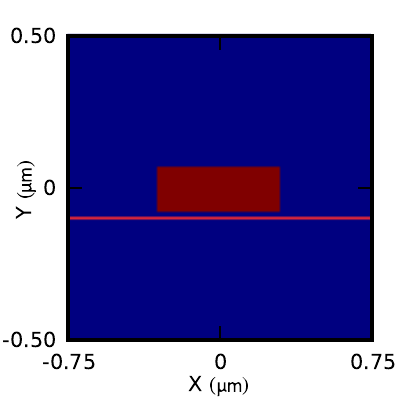}
		} &
		\mbox{
		  \includegraphics{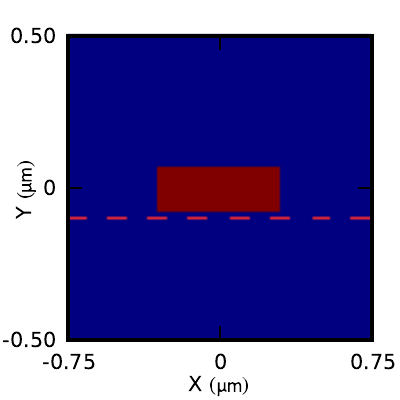}
		} &
		\mbox{
		  \includegraphics{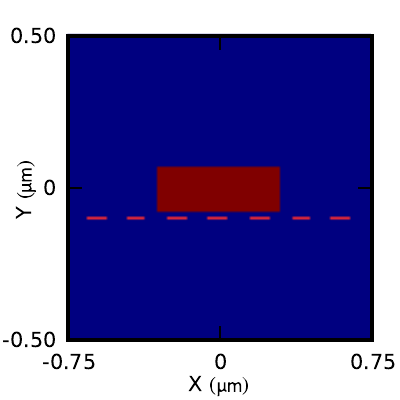}
		}
		\\
		\hspace{0.3cm}\bf{(a)} & \hspace{0.3cm}\bf{(b)} & \hspace{0.3cm}\bf{(c)}
		\\
		\mbox{
		  \includegraphics{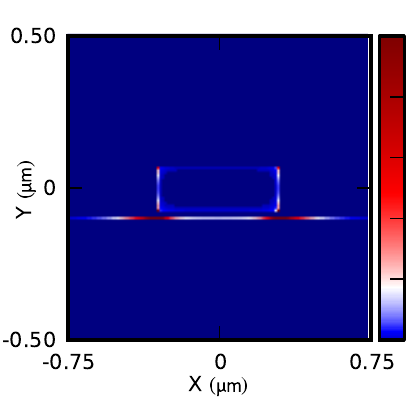}
		} &
		\mbox{
		  \includegraphics{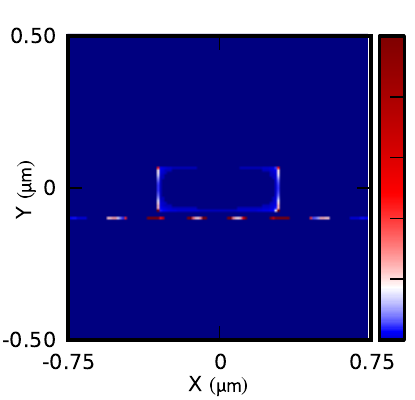}
		} &
		\mbox{
		  \includegraphics{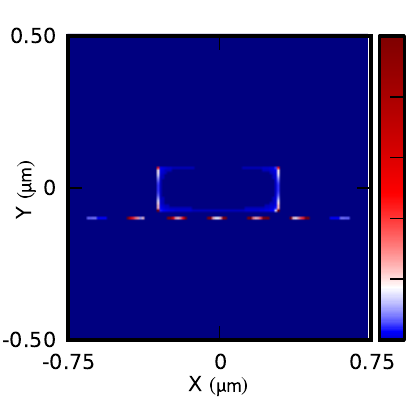}
		} \\
		\hspace{0.3cm}\bf{(d)} & \hspace{0.3cm}\bf{(e)} & \hspace{0.3cm}\bf{(f)}
	\end{tabular}
\end{center}
\caption{\label{fig:loss_structures} \textbf{Absorption calculations. }
Three simulated geometries (a - c) and their mode-loss distributions ($Im(\epsilon)E^2$, d - f) to determine the optimum configuration for plasmon absorbtion. (a, d) SSPD orthogonal to waveguide. This geometry gives an absorbtion of $61.3 \%$ in the NbN film, but has a duty cycle of only $50 \%$. (b, e) SSPD parallel to waveguide, waveguide edges at center of NbN stripe, absorbtion in NbN film: $66.8 \%$. (c, f) SSPD parallel to waveguide, waveguide edges at center of gap between NbN stripes, absorbtion in NbN film: $51.9 \%$.
}
\end{figure}

\subsection{Absorption calculations}
To optimize the fraction of optical power absorbed in the SSPD (NbN, $\epsilon = -16.96 + 13.05i$) versus losses in the plasmonic waveguide, we calculate the eigenmodes of several geometries in \reffig{fig:loss_structures}a-c. The aborbtion loss for these structures is given by $\int{Im(\epsilon)E^2dA}$ and shown in \reffig{fig:loss_structures}d-f. We performed these simulations for different waveguide widths, but only show the results of a width of $600 \unit{nm}$. We conclude that the optimal geometry is the one with the SSPD running parallel to the waveguide and having the waveguide edges right above the NbN meander stripes.

\newpage
\bibliographystyle{ieeetr}

\begin{thebibliography}{1}

\bibitem{fedrizzi2007wavelength}
A.~Fedrizzi, T.~Herbst, A.~Poppe, T.~Jennewein, and A.~Zeilinger, ``A
  wavelength-tunable fiber-coupled source of narrowband entangled photons,''
  {\em Opt. Express}, vol.~15, no.~23, pp.~15377--15386, 2007.

\bibitem{ou1989fourth}
Z.~Y. Ou, E.~C. Gage, B.~E. Magill, and L.~Mandel, ``Fourth-order interference
  technique for determining the coherence time of a light beam,'' {\em JOSA B},
  vol.~6, no.~1, pp.~100--103, 1989.

\bibitem{joobeur1994spatiotemporal}
A.~Joobeur, B.~Saleh, and M.~Teich, ``Spatiotemporal coherence properties of
  entangled light beams generated by parametric down-conversion,'' {\em Phys.
  Rev. A}, vol.~50, no.~4, p.~3349, 1994.

\bibitem{mandel1986non}
L.~Mandel, ``Non-classical states of the electromagnetic field,'' {\em Phys.
  Scr. T}, vol.~12, pp.~34--42, 1986.

\bibitem{jung2007theoretical}
J.~Jung, T.~S{\o}ndergaard, and S.~I. Bozhevolnyi, ``Theoretical analysis of
  square surface plasmon-polariton waveguides for long-range
  polarization-independent waveguiding,'' {\em Phys. Rev. B}, vol.~76, no.~3,
  p.~035434, 2007.

\bibitem{agrawal2001fiber}
G.~P. Agrawal, {\em Nonlinear fiber optics}.
\newblock Academic Press, 2001.

\end{thebibliography}

\end{document}